# Characteristic time scales of UV and IR auroral emissions at Jupiter and Saturn and their possible observable effects

Chihiro Tao[1], Sarah V. Badman[1], and Masaki Fujimoto[1]

[1]*ISAS/JAXA, Yoshinodai 3-1-1, Chuo-ku, Sagamihara, 252-5210, Japan*
*Email : tao@stp.isas.jaxa.jp*

Different ultraviolet (UV) and infrared (IR) auroral features have been observed at Jupiter and Saturn. Using models related to UV and IR auroral emissions, we estimate the characteristic time scales for the emissions, and evaluate whether the observed differences between UV and IR emissions can be understood by the differences in the emission time scales. Based on the model results, the UV aurora at Jupiter and Saturn is directly related to excitation by auroral electrons that impact molecular $H_2$, occurring over a time scale of $10^{-2}$ sec. The IR auroral emission involves several time scales: while the auroral ionization process and IR transitions occur over < $10^{-2}$ sec, the time scale for ion chemistry is much longer at $10^{-2}$–$10^{4}$ sec. Associated atmospheric phenomena such as temperature variations and circulation are effective over time scales of > $10^{4}$ sec. That is, for events that have a time scale of ~100 sec, the ion chemistry, present in the IR but absent in the UV emission process, could play a key role in producing a different features at the two wavelengths. Applying these results to the observed Jovian polar UV intensification events and the Io footprint aurora indicates that whether the IR intensity varies in correlation with the UV or not depends on the number flux of electrons and their characteristic energy.

## 1 Introduction

Outer planetary aurorae are emitted across a wide range of wavelengths which enables remote sensing of various physical parameters of the planetary environment. Ultraviolet (UV) and infrared (IR) wavelengths are the most commonly observed. The UV emissions come from hydrogen directly excited by auroral electrons. The IR wavelengths are emitted from thermally excited $H_3^+$ which is produced from the ionization of atmospheric hydrogen by auroral electrons and solar EUV.

Focusing on the similarities and differences between UV and IR emission mechanisms, comparisons between them have been made based on statistical features and near simultaneous observations (Clarke et al., 2004). For Jupiter, UV and IR images separated by two minutes show different emission intensities along the main oval and in the polar region, a UV-only low latitude extension, and different Io footprint intensity compared with the main oval (Clarke et al., 2004). The location of Saturn's main oval is statistically similar in the UV and IR (Badman et al., 2011). Stallard et al. (2008) reported an IR enhancement over a large area in Saturn's polar region, which is located at higher latitude than the main oval associated with the open-closed field line boundary. Cassini observations provide unique simultaneous observation of UV and IR aurorae, showing similar emission in the main oval, but different emissions in other lower or higher latitude regions (Melin et al., 2011).

Regarding emissions from the polar regions of the two planets, Tao et al. (2011) focused on the Saturnian IR-only polar emission reported by Stallard et al. (2008) to constrain the possible causal atmospheric temperature and auroral electron properties using an emission model. Since IR emission from Saturn is strongly dependent on temperature, several 10s or a few 100s K heating could cause these IR polar emissions. Time variations of the UV aurora are observed in Jupiter's polar region over various time scales from several seconds (Waite et al., 2001) to several days related with magnetotail reconnection (e.g., Grodent et al., 2004). Since different emission mechanisms contain different time scales, it is important to consider time variations in comparisons between the UV and IR. In other words, comparative UV-IR studies would tell us more about the underlying mechanisms that produce the auroral features seen at the outer planets.

The purpose of this study is to investigate how UV and IR emissions vary with time, and what information can be deduced from observations. Here we report mainly the former topic as follows. In Section 2 we estimate several characteristic time scales related with the emissions. The time variation of IR emission intensity with incident electron energy and flux is characterized in Section 3. In Section 4 our emission models are applied to the polar events and Io footprint aurora described above, to explain the similarities and differences observed between the UV and IR emissions. Finally Section 5 summarizes this study.

## 2 Time scale estimation

Time scales related with emission processes provide an important framework for not only understanding the observed auroral time variations but also modeling studies. Since it requires a large simulation load to include all time scales from auroral electron collisions to atmospheric dynamics simultaneously within a model, it is necessary to judge whether time variation should be solved or considered as a constant, by comparison of the appropriate time scales with the time variation of each phenomenon of interest.

### 2.1 Overview

Fig. 1 shows the processes from auroral electron precipitation to UV and IR emissions with their characteristic time scales, which are described in the following sections. The UV aurora is emitted from electron-excited hydrogen when it de-excites to its ground state. Auroral electrons also ionize molecular hydrogen, which can undergo various chemical reactions to produce ions including $H_3^+$. Following collisions with background $H_2$ under high thermospheric temperature, $H_3^+$ is excited vibrationally. Some excited $H_3^+$ ions de-excite by IR emission to generate the aurora of interest in this study. For more details of these processes the reader is referred to Tao et al. (2011).

### 2.2 Auroral electron precipitation

The auroral precipitation process is evaluated using a Monte Carlo simulation model (Hiraki and Tao, 2008), which solves auroral electron precipitation into the Jovian $H_2$ atmosphere including elastic, ionization, excitation, vibration, and rotational collisions. Fig.2a shows altitude profiles of collision rates time-integrated after electrons with initial energy of 10 keV were incident at the upper boundary, placed at 2500 km for Jupiter. The colour-coding of the lines represents the different collisional processes, described by Hiraki and Tao (2008), as labeled in Fig. 2b. 0.025 sec after incidence of the electrons, the ionization and B/C excitation collisions, shown in red and orange, respectively, are dominant along their trajectory, because of their large cross section in the > 1 keV energy range. As the incident electrons reduce their energy and secondary electrons are produced through the ionization process, the number of low energy electrons increases. This is clearly seen as the increase of rotation and vibration collisional processes (black and purple lines) evident at low altitudes after 0.05 sec, followed by an increase after 1 sec at high altitudes where the collision frequency is low. Altitude profiles of the collision time scales are shown in Fig. 2b. As seen above, the time scales of ionization and excitation are $\sim 2\times 10^{-2}$ sec over the altitudes studied, which is similar to and smaller than the vibrational time scales at low and high altitudes, respectively.

### 2.3 Ion chemistry

Our model solves a simplified set of neutral-ion chemical reactions (see Tao et al. (2011)) for 13 ions ($H^+$, $H_2^+$, $H_3^+$, $H_2O^+$, $H_3O^+$, $CH_3^+$, $CH_4^+$, $CH_5^+$, $C_2H_2^+$, $C_2H_3^+$, $C_2H_5^+$, $C_3H_n^+$, $C_4H_n^+$, where the latter two symbols represent classes of ions) and six fixed neutral species ($H_2$, H, $CH_4$, $C_2H_2$, $C_2H_4$, $H_2O$) taking into account the ambipolar diffusions of $H^+$ and $H_3^+$. Fig.3 shows altitude profiles of the characteristic time scales derived from the $H_3^+$ density multiplied by its production and loss rates in the steady state. At low altitude, <500 km at Jupiter (Fig.3a) and <1000 km at Saturn (Fig.3b), the dominant loss process, i.e., the process with the shortest timescales, is recombination with hydrocarbons. The associated time scales for Jupiter and Saturn are <100 sec and <10 sec, respectively. The presence of $H_2O$ at Saturn leads to reactions with water group molecules being the dominant loss process at 1000–1500 km altitudes. At higher altitudes, the electron recombination loss process is dominant with a time scale of 100–1000 sec. Ambipolar diffusion becomes the main loss process at >2000 km at Jupiter with a time scale of >1000 sec.

## 2.4 Vibrational equilibrium

The distribution of $H_3^+$ among its vibrational levels follows local thermodynamic equilibrium (LTE) when collisions with $H_2$ are frequent enough to recover departures from LTE resulting from de-excitation through IR emission. The departure from LTE thus increases with decreasing $H_2$ density at high altitudes. The reduction of vibrationally excited $H_3^+$ decreases the IR emission rate. Using the equilibrium state of vibrational levels between collision with $H_2$ and IR transitions (Tao et al., 2011), production and loss time scales for vibrational levels of $\nu_2(1)$ are estimated and shown in Fig. 4. Collisional production and loss processes are dominant at <1000 km for Jupiter (Fig. 4a) and <2000 km for Saturn (Fig. 4b); the time scale of these increases with altitude as the $H_2$ density decreases. For higher altitudes, loss by IR emission becomes dominant and determines the time scale to be $3.8\times10^{-3}$ sec. Departure from LTE becomes significant at these altitudes (where the timescale of loss by emission becomes shorter than that of collisional production) to ~3000 km. The altitude profiles of the $2\nu_2(0)$ and $2\nu_2(2)$ vibrational states (Figs.4c and 4e for Jupiter and Figs. 4d and 4f for Saturn, respectively), which are related to ~2 μm IR emissions, are similar to the $\nu_2(1)$ profiles.

## 2.5 UV and IR emission

The time scales for UV and IR emissions are estimated as the reciprocal of the probability of vibrational transition, i.e., the Einstein coefficient. Although the time scales of transitions from the $H_2$ B and C states to the ground state, which cover a large part of the UV emission, vary across the range $10^{-9}$–$10^{3}$ sec, the wavelength-integrated intensity varies over ~$10^{-8}$ sec. The emission time scales of the main IR lines detected by ground-based observations in the 2 μm (Raynaud et al., 2004) and 4 μm (Lystrup et al., 2008) wavelength ranges are $10^{-3}$–$10^{-2}$ sec.

## 2.6 Dynamic transport and energetics

A one-dimensional local model provides a good approximation for events with time scales smaller than those of dynamic transport or temperature changes. Referring to the quasi-steady state thermosphere models for Jupiter (Tao et al., 2009; Bougher et al., 2005) and Saturn (Smith et al., 2007; Müller-Wordarg et al., 2006), Table 1 lists estimated time scales in the region around the main oval located at ~75° latitude and at the $H_3^+$ peak altitude for the case of 10 keV electron precipitation, i.e., 500 km for Jupiter and 1500 km for Saturn. We find time scales by transport and $H_2$ diffusion are $10^4$–$10^5$ sec and $10^6$–$10^8$ sec, respectively.

Considering the effects of heating, characteristic time scales are estimated as $c_p/Q$ (sec/K), where $c_p$ (J/K kg) is the $H_2$ heat capacity and $Q$ (W/kg) is the heating or cooling rate. We consider the following processes for heating: the sum of the meridional and vertical advection terms (adv_h), adiabatic heating/cooling (adi), the work done by viscosity ($F_{visV}$), the work done by ion drag ($F_{ionV}$), heat conduction ($Q_{con}$), auroral particle heating ($Q_{aur}$), solar EUV heating ($Q_{sEUV}$), IR cooling ($Q_{IR}$), wave heating ($Q_{wave}$), and Joule heating ($Q_J$) (Tao et al., 2009). For Jupiter, Fig. 5 shows the altitude profiles of the time scales associated with these processes averaged over 65–80° latitude. The shortest timescales overall are $10^3$–$10^4$ sec/K at different altitudes. The time scales become small with decreasing $H_2$ density at high altitudes. For Saturn, we take heating and cooling rates of 3–10 W/kg (supplementary information of Smith et al., 2007) providing time scales of $10^3$ sec/K.

Note that the heating and cooling effects vary by up to a few orders of magnitude, depending on magnetospheric and atmospheric conditions. In addition, heating of $H_3^+$ itself and related latitudinal transport should ideally be taken into consideration, while here we assume for simplicity that enough collisional interaction occurs to result in similar temperatures of $H_2$ and $H_3^+$.

## 2.7 Summary of relative time scale analysis

The above results indicate that $H_2$ excited by auroral electrons immediately produces UV emission. For studies of the IR emission time variation, auroral electron ionization and IR emission transfer should be considered for changes occurring over $<10^{-2}$ sec (as shown in Fig. 2). Atmospheric dynamics and temperature changes occur over $>10^4$ sec (Fig. 5). The ion chemistry time scale is important in the intermediate range: $10^{-2}$–$10^4$ sec (Fig. 3).

Later in this paper, we have chosen to study two temporally varying Jovian auroral processes: the Io footprint emission and the polar auroral intensifications. The timescale for the Io event is 404 sec which corresponds to the time taken for the Io footprint main spot to pass over a point on the planet's surface (the detailed description is in Section 4.2). The polar intensifications occur over timescales of $10^2$–$10^3$ sec. At these timescales, the ion chemistry in the IR emission process would play a key role and there is a reasonable expectation that the associated IR emission would behave differently from the UV emission. Hereafter we thus consider time variations of IR intensity related to ion chemistry under constant temperature and assuming instantaneous vibrational transitions.

# 3 IR time variation

In Fig. 6a the intensity from the Q(1,0) emission line at Jupiter is shown as a function of incident electron energy and flux, scaled to the emission produced from a flux 0.15 μA/m$^2$ electrons with energies of 10 keV (after Tao et al., 2011). There are many possible changes in the incident electron parameters which can cause a factor of 2 variation in emission intensity. We test the time variation of three of the possibilities shown by the thick lines, considering either energy or flux variation. The time variations of the integrated IR intensity for the three cases which enhance IR emission are shown in Fig. 6b. When the electron energy was increased from 1.18 to 10 keV (dot-dashed line) the IR intensity showed the quickest variation. This was followed by the case where the electron flux was increased from 36 to 150 nA/m$^2$ (solid line), while the case where the electron energy was decreased from 95 to 10 keV (dashed line) showed the slowest change in IR intensity. The energy increase (decrease) shifts the emission altitudes lower (higher) while the flux change provides changes in emission intensity at the same altitudes (Fig. 6c). As seen in Fig. 3, the ion chemical time scale is smaller at low altitude. Therefore the different time variations of these three passes are understood in terms of the different time scales across altitude. In Figs. 6d and 6e the corresponding profiles for the changes which decrease the IR intensity are shown. The time variation is faster in the case where the electron flux was decreased than in the cases where the electron energy was varied. This slower intensity variation related to changing the electron energy would be caused by the IR emission from new altitudes according to the electron energy, i.e., >1000 km for the lower energy electrons and <500 km for the higher energy electrons, as shown in Fig. 6e.

## 4. Estimation for time-variable events

### 4.1 Jupiter polar region

Bonfond et al. (2011) reported variations in the UV intensity of Jupiter's polar region with a timescale of minutes. Here we estimate how the IR emission would accompany this temporally-varying UV emission using the time-variable emission model. Variation in the incident electron parameters is assumed to produce the variation in emitted power observed on September 11, 2009 (Bonfond et al., 2011). Figs. 7a and 7b show the three incident electron models considered. It is assumed that variation of either electron energy (dashed and dotted lines) or number flux (solid line) is responsible for the observed UV intensity variation. These are log-scaled profiles following the emission power profile in Bonfond et al. (2011). An upper limit on the electron energy variations is set at 50 keV because UV absorption by hydrocarbons becomes effective at higher energies and the relationship between an electron energy increase and a UV intensity increase becomes so complex that it is beyond the simple test performed here. The estimated IR time variations in Fig. 7c are plotted after a 100 sec running average is subtracted. Vertical displacements (as labeled) are made so that the variations in each case can be seen clearly. The IR variations associated with the electron flux variation (solid lines in Fig. 7c) and the energy variation in the low energy range of <10 keV (dotted line) are correlated with the UV variations with a time lag of ~100 sec. On the other hand, the IR intensity due to the energy variation at a higher energy >10 keV (dot-dashed line) shows different variations and is inversely correlated with the UV at around 1100 sec. These different IR variations associated with different electron energies are explained by the fact that IR intensity decreases with increasing energy above 10 keV. As shown in Fig. 6a, >10 keV electrons precipitate to lower altitudes and the emission process proceeds in an environment where the temperature is lower and where $H_3^+$ recombines with hydrocarbons. The similarities and differences between the UV and IR are largely affected by the electron energies.

### 4.2 Io footprint aurora

Io's footprint aurora is composed of multiple spots. From the location of the spots, they are supposed to be produced by Alfven waves reflected within the Io torus (e.g., Gérard et al., 2006) and electron beams reflected at the ionosphere (Bonfond et al., 2008). Clarke et al. (2004) mention that in the near-simultaneous UV and IR images they have shown the Io footprint aurorae have UV intensity comparable with the main oval while the IR intensity is lower.

Here we focus on the Io main spot aurora and estimate how the IR emission intensity varies due to time variation, electron energy, and background temperature. The longitudinal width of the Io footprint spot is ~3° (Bonfond et al., 2008). Using the corotation velocity at Io's orbit $V_{cor} = 2\pi \times 5.9 \times 71500\text{km} / 35700\text{sec} = 74.2$ km/s and Io's orbital velocity $V_{Io} = 2\pi \times 5.9 \times 71500\text{km} / (42 \times 3600) = 17.5$ km/s, The pass time of Io's footprint spot at a certain longitude becomes $2\pi \times 71500\text{km} \times \cos(65°) \times (3°/360°) \times (5.9/\cos(65°)) / (V_{cor} – V_{Io}) = 404$ sec. Electron precipitation would continue at Io's footprint during this time. For the Io case, the characteristic energy is taken as the parameter to survey. While the energy is varied, the incident electron flux is set to produce the same UV intensity as at the main oval which we consider to be caused by electrons with characteristic energy of 10 keV and flux of 0.15 $\mu A/m^2$. The atmospheric temperature can also be different between the main oval and the Io footprint. Since the Io footprint is located at ~10° lower latitude than the main oval where the auroral energy input dominates, the background thermospheric temperature at the footprint would be less than that in the main oval. The IR intensity variation for 10 keV incident electrons is shown in Fig. 8a. For a background temperature of 1000 K, the IR intensity increases during precipitation and decreases after precipitation has ceased (solid line). The UV intensity enhances only during the electron precipitation, i.e., 0-404 sec. Continuous precipitation at temperatures of 1000 and 1200 K are shown by the dot-dashed and dashed lines. The quasi-steady states obtained at >500 sec are considered to represent the main oval situation. Fig. 8b shows the maximum IR intensity obtained for the Io footprint case (the solid line in Fig.8a) as a function of electron energy. The two main oval conditions (1000 K and 1200 K) are depicted by the two symbols. There is a peak at a few keV that has the best combination of a moderate IR emission efficiency and a moderate electron number flux (note that electron number flux is variable with energy while Fig. 6a shows variations under constant flux conditions). When compared at the same electron energy, the maximum intensity of the representative main oval conditions are higher than the Io footprint conditions because the max intensities are not attained until ~500 sec, i.e., longer than the Io footprint pass time. The difference in max intensity between the two 1000 K and 1200 K main oval conditions reflects the temperature-dependence of the IR emission. How these two factors (time scale and temperature) that are different between the main oval and the Io footprint determine the IR intensity ratio is plotted in Fig. 8c. The ratios caused by time variation is electron precipitation (crosses) and by temperature in addition to the time variation (asterisks) are plotted as a function of electron energy. These correspond to the ratios of the solid line to the cross and the asterisk, respectively, in Fig. 8b. The time variation effect both increases and decreases the intensity ratio depending on electron energy, while including the temperature effect further reduces the intensity ratio by 15–87%. If the Io footprint is caused by electrons with a few keV (e.g., Bonfond, 2010), a temperature reduction and/or more localized electron precipitation (i.e., shorter precipitation time in the model) should be considered to obtain the reduction in IR intensity of the Io footprint reported by Clarke et al. (2004).

## 5. Summary

We have estimated the time scales of processes related to UV and IR emissions from Jupiter and Saturn using atmospheric models, with the following results:
(i) The time scale for ionization and excitation by auroral electrons is ~ $10^{-2}$ sec.
(ii) The ion chemistry time scale increases with altitude from < 1 to $10^4$ sec.
(iii) $H_3^+$ vibrational levels have lifetimes of < $10^{-2}$ sec.
(iv) The time scale of the main UV and IR emission lines are $10^{-8}$ and $10^{-2}$ sec, respectively.
(v) Transport by atmospheric dynamics requires >$10^5$ sec, and variations in temperature from heating and cooling effects take ~$10^3$ sec/K.
(vi) IR intensity variations observed over several minutes will be affected by chemical time scales varying with altitude.

Applying the emission model to Jupiter's polar emission and the Io footprint aurora, the IR intensity variation is estimated as follows:
(vii) IR variations due to electron flux or <10 keV energy modulations are correlated with UV variations with ~100 sec time lag. IR variations due to higher energy >10 keV modulations vary differently and are sometimes inversely correlated with the UV.
(viii) The IR intensity of the localized Io footprint spot relative to the main oval intensity is reduced by 15–87% by the combination of time variations of electron precipitation and lower temperature.

Although (vii) and (viii) are preliminary results to be tested using more auroral electron energy and flux cases, and to be evaluated by comparison with IR and UV statistical observations, this model is a useful tool for investigating the observed auroral emissions.

Table 1. Time scales for atmospheric dynamics

| | Jupiter | Saturn |
|---|---|---|
| $H_2$ horizontal transport<br>$\tau$ = (oval width)/ $v_{latitude}$ | (1000km)/(10m/s)<br>= $10^5$ sec | (2000km)/(100m/s)<br>= $2\times10^4$ sec |
| $H_2$ vertical transport<br>$\tau$ = (scale height)/ $v_{altitude}$ | (100km)/(1m/s)<br>= $10^5$ sec | (100km)/(1m/s)<br>= $10^5$ sec |
| $H_2$ vertical diffusion<br>$\tau$ = (scale height)$^2$/(dif.coef.) | (100km)$^2$/(100m$^2$/s)<br>= $10^8$ sec | (100km)$^2$/($10^{3-4}$m$^2$/s)<br>= $10^{6-7}$ sec |

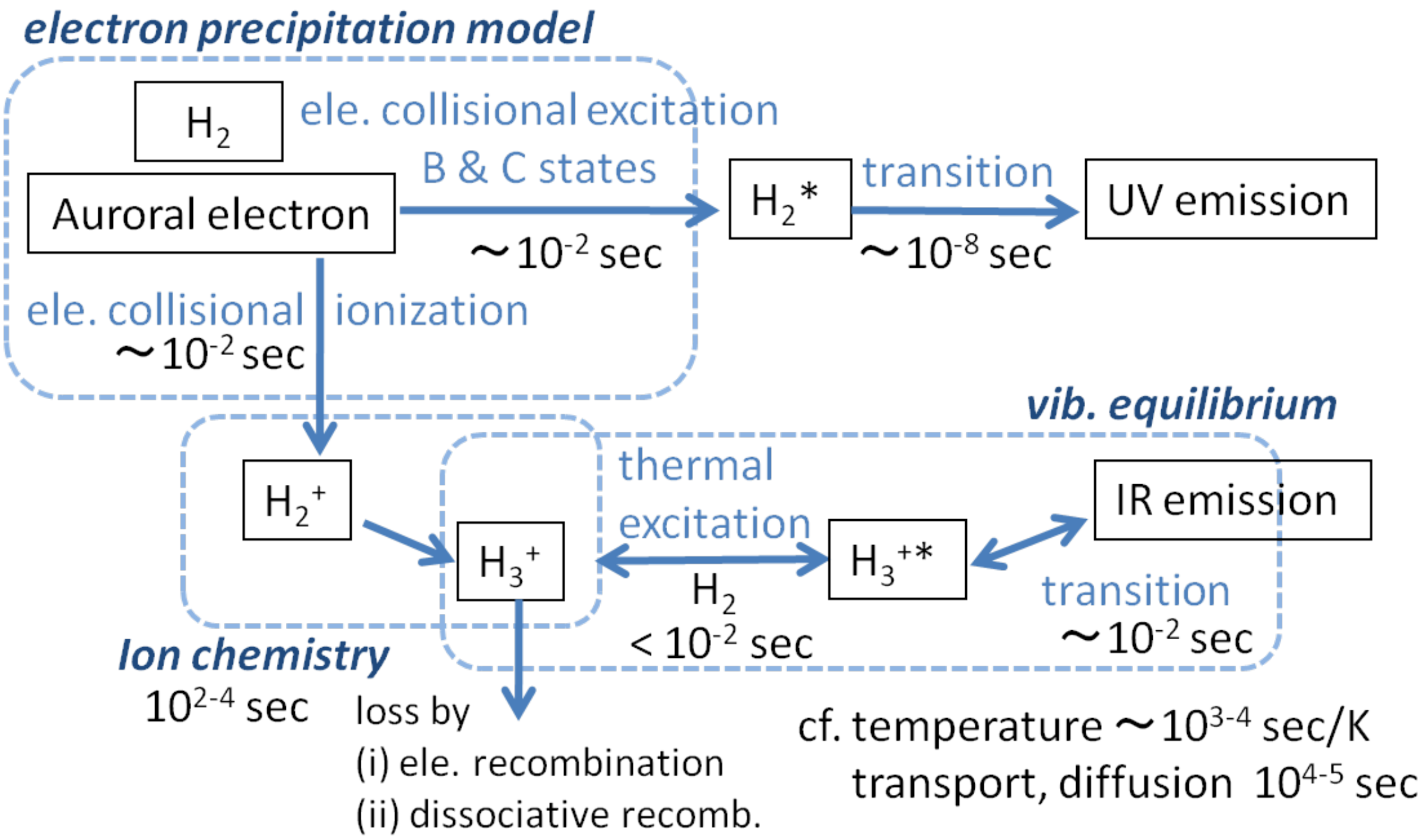

Fig.1. Flowchart of UV and IR auroral emissions. See details in the text.

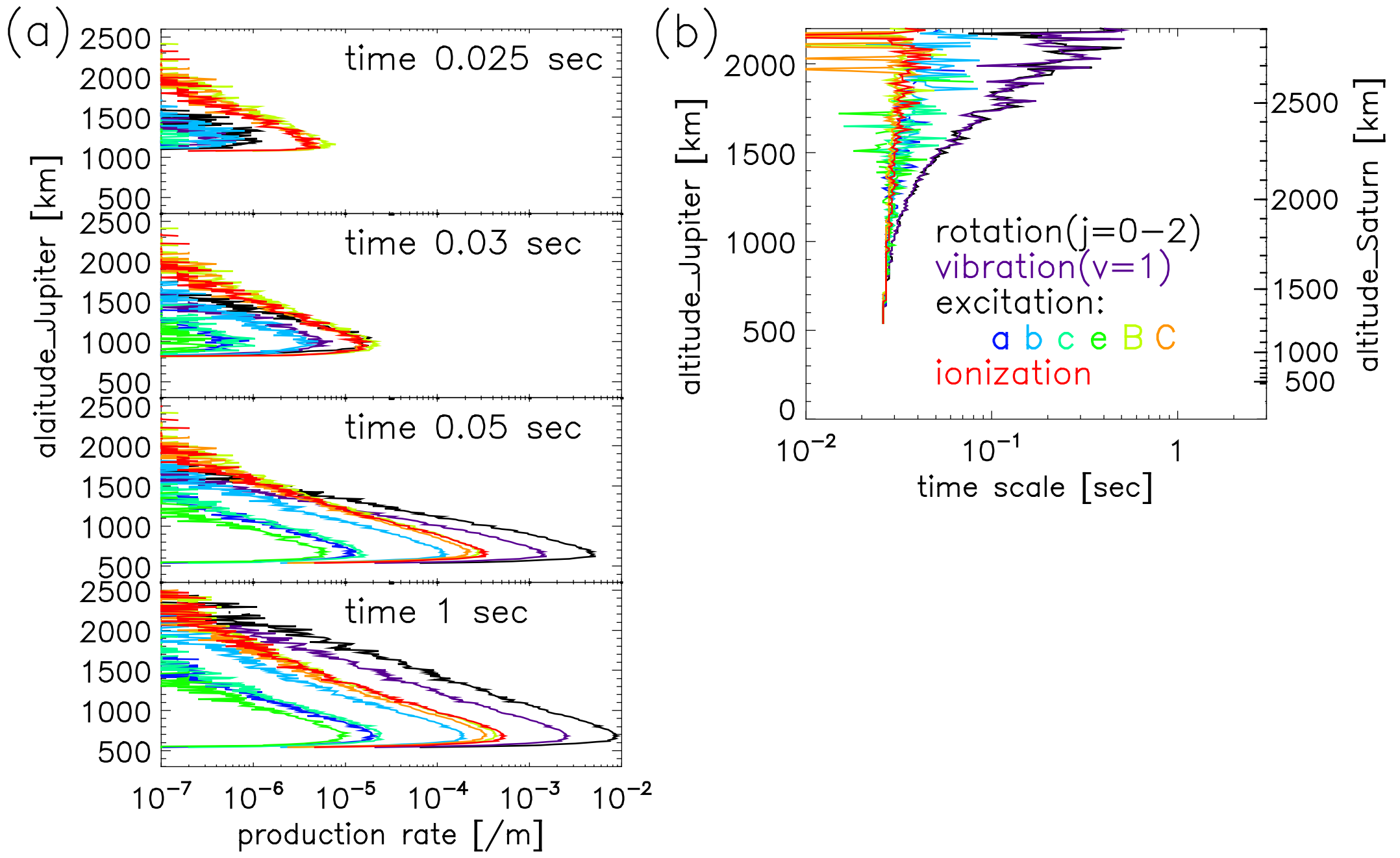

Fig.2 (a) Time development of collision rates per electron with initial energy of 10 keV. Panels from top to bottom show time integration after electron incident until 0.025, 0.03, 0.05, and 1 sec, respectively. The colour-coding of the lines represents different collision processes as labeled in (b). (b) Altitude profiles of time scales of collision processes. The corresponding Saturn atmospheric altitudes are shown in the left axis.

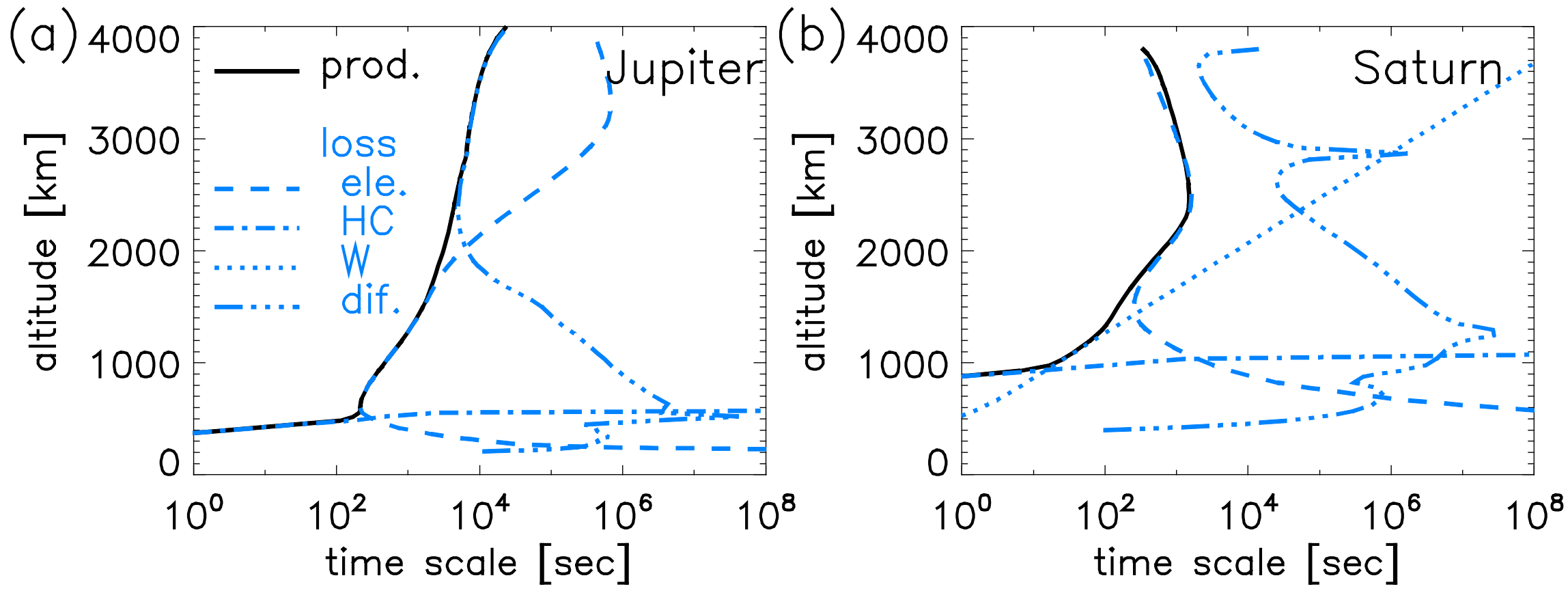

Fig.3 Altitude profiles of chemical time scales for $H_3^+$ production (solid line) and loss processes for (a) Jupiter and (b) Saturn. The considered $H_3^+$ loss processes are recombination with electrons (dashed line), hydrocarbons (dot-dashed), and water-group molecules for the Saturn case (dotted), and ambipolar diffusion (triple-dot-dashed).

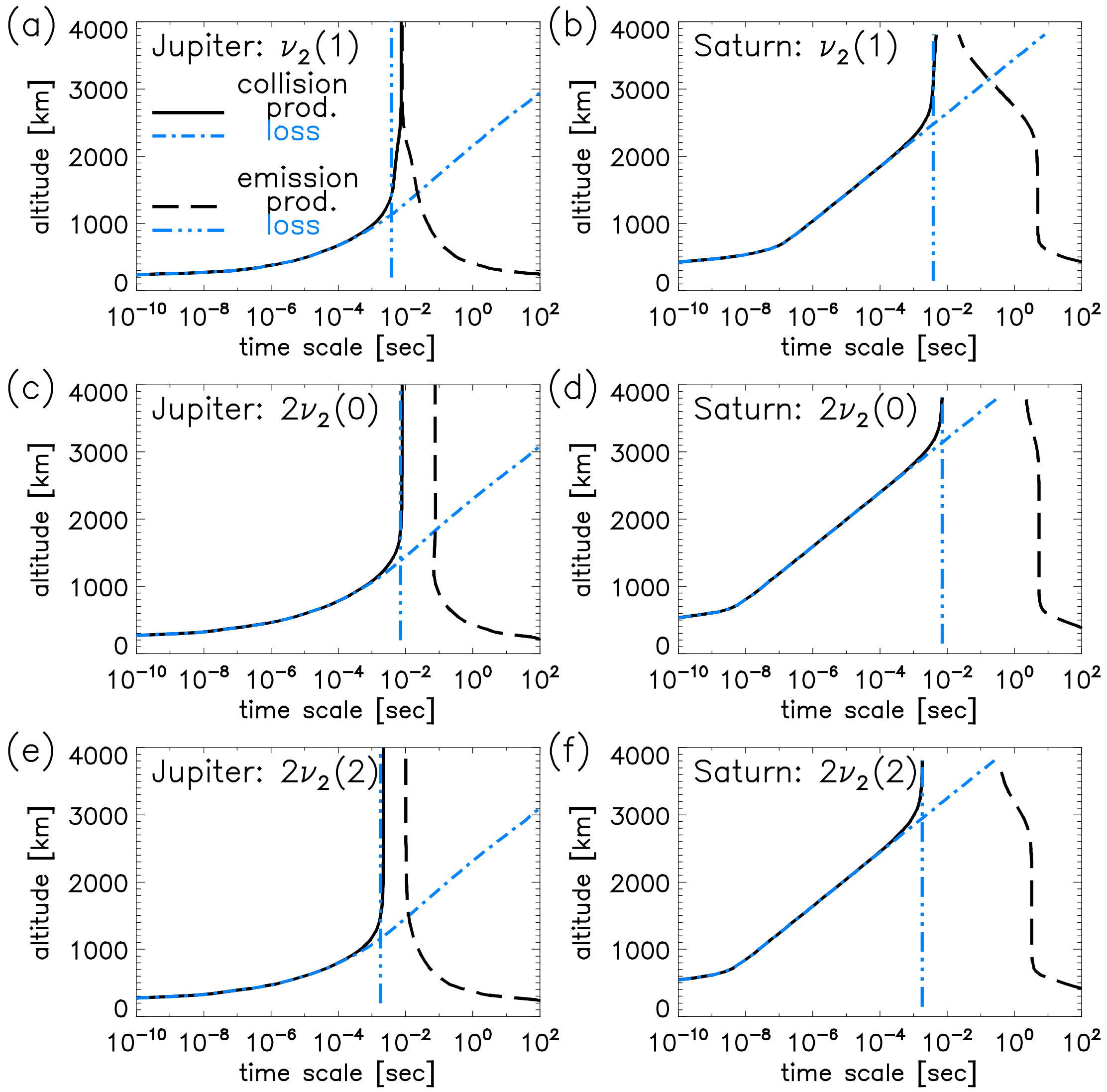

Fig.4 Altitude profiles of time scales for $H_3^+$ vibrational level (a) $v_2(1)$ for Jupiter and (b) Saturn, (c) $2v_2(0)$ for Jupiter and (d) Saturn, and (e) $2v_2(2)$ for Jupiter and (f) Saturn. Black lines show production by collisions (solid line) and emission transitions (dashed), and blue/grey lines show loss by collisions (dot-dashed) and emission transitions (triple-dot-dashed).

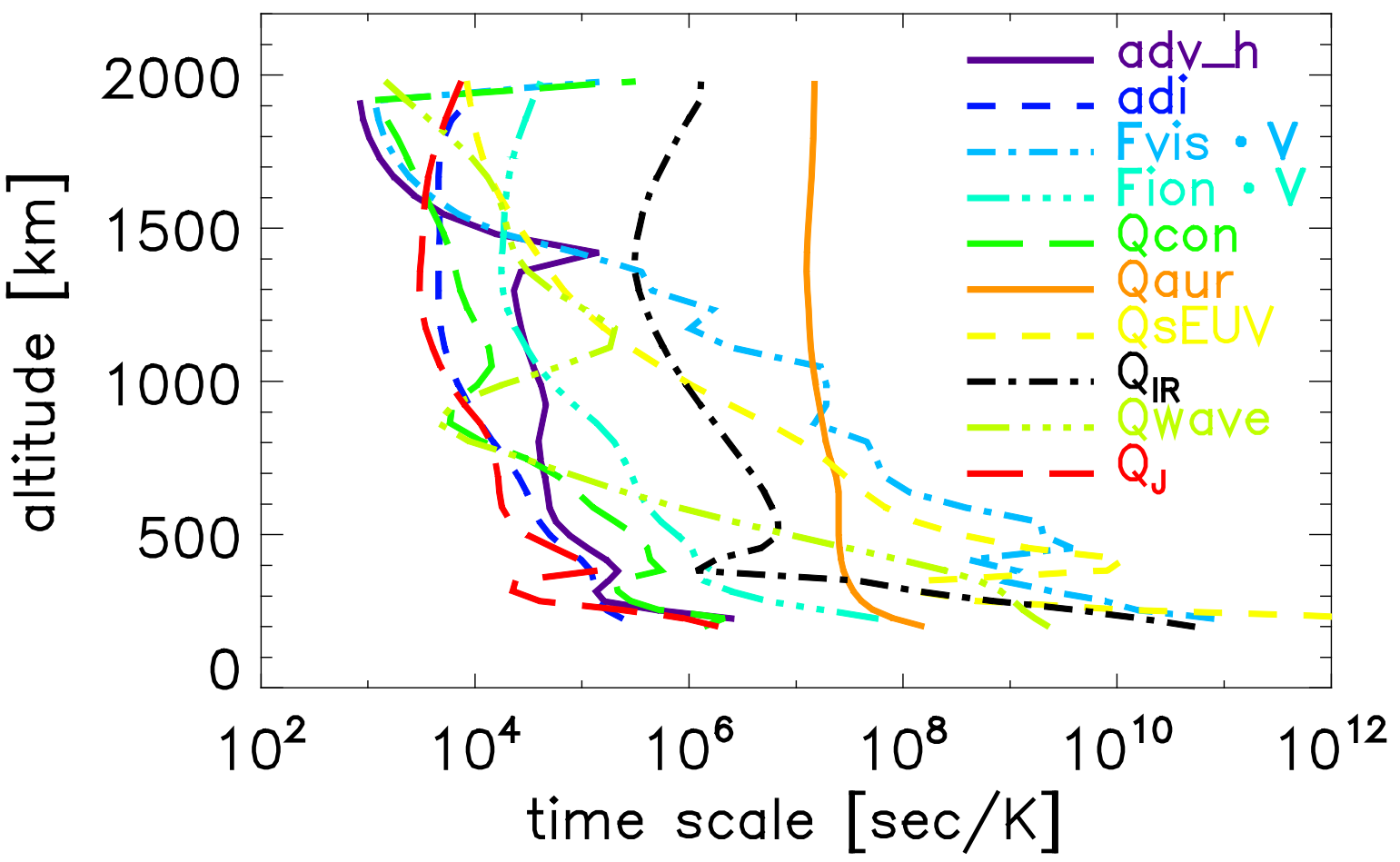

Fig.5 Altitude profiles of characteristic time scales of Jovian temperature variations by each process labeled on the right. See details in the text.

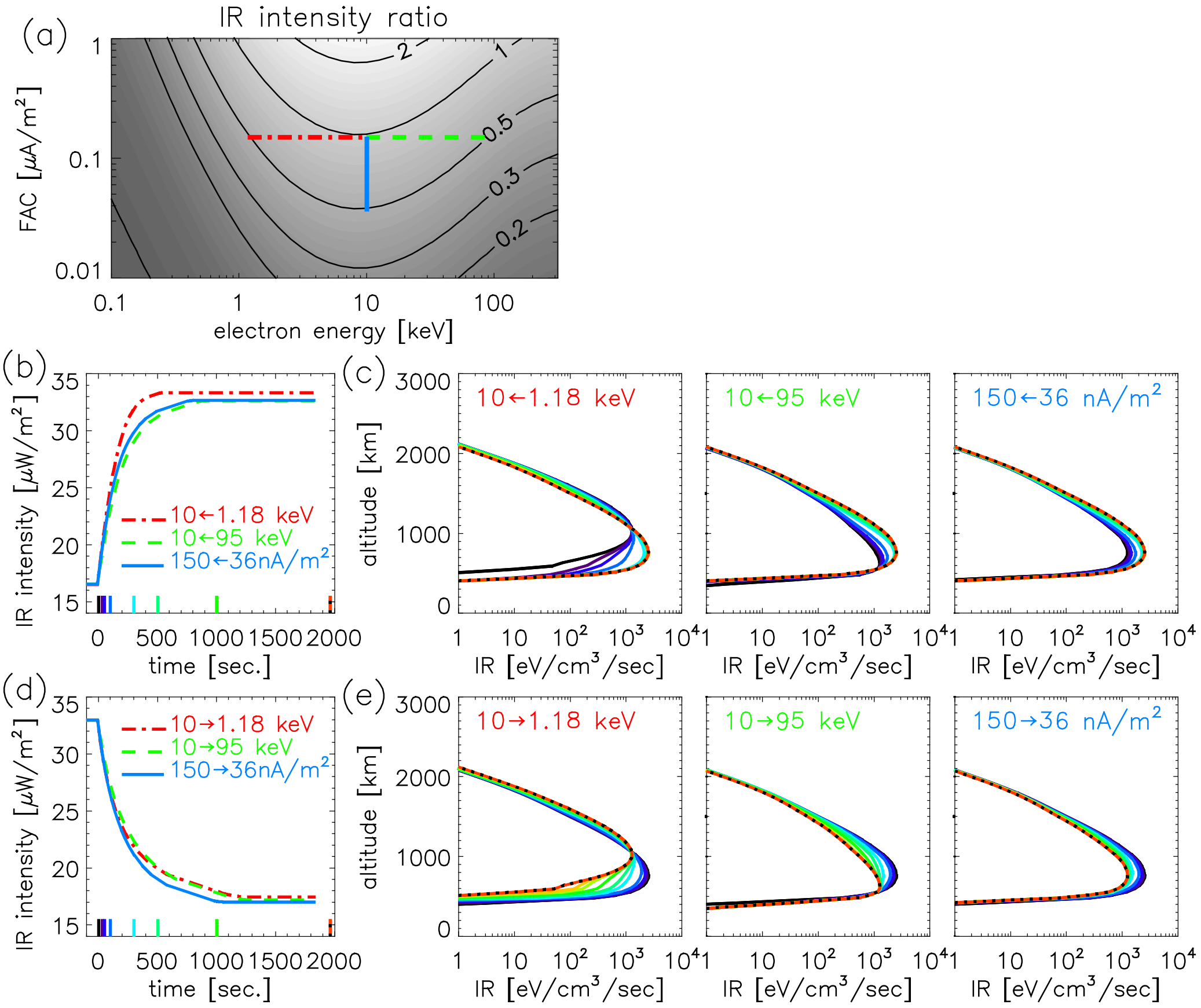

Fig.6 (a) IR emission intensities normalized to the case with electron energy of 10 keV and flux of 0.15 μAm², time variation of altitude-integrated IR emission for (b) increasing and (d) decreasing cases along thick lines in (a). (c) shows time variation of altitude profiles of three cases in (b) with the corresponding time is shown by vertical lines in low part of (b), and (e) is those of (d).

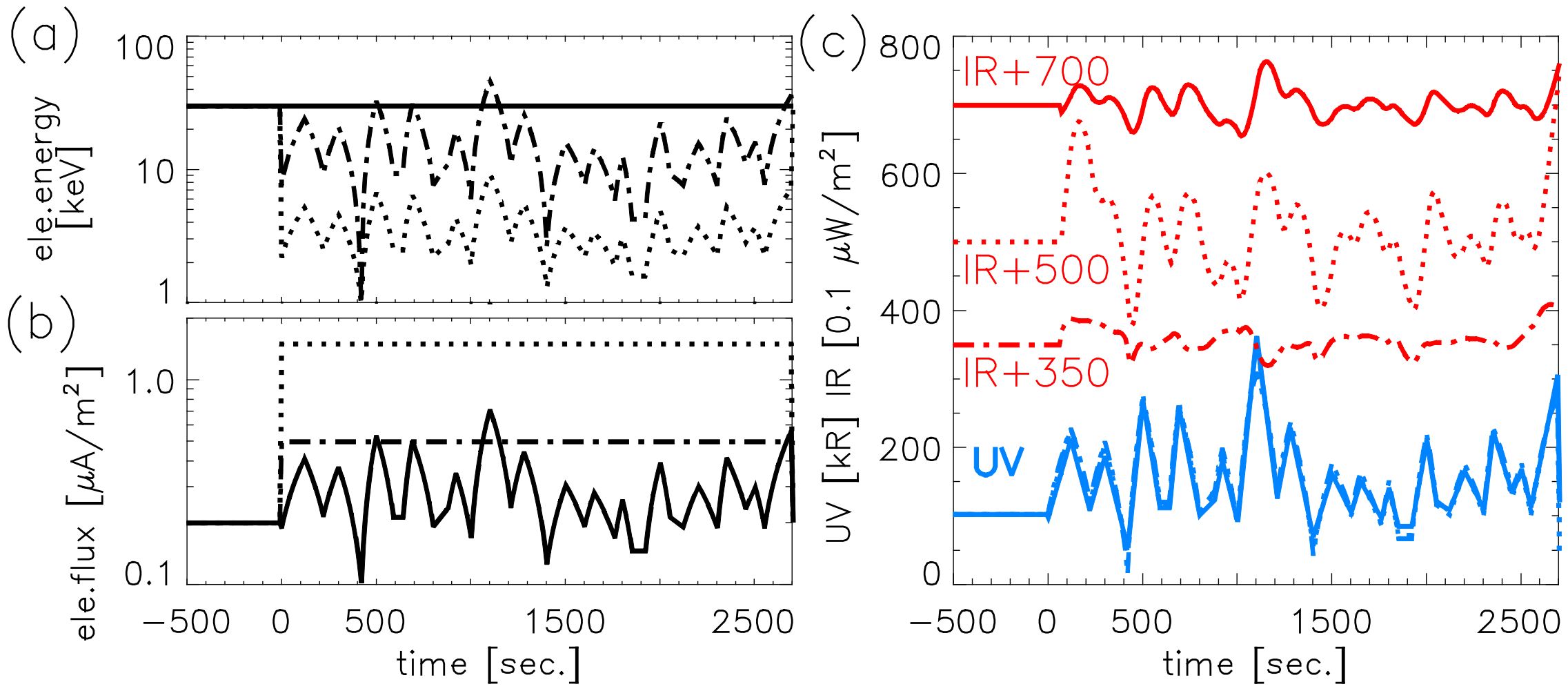

Fig.7 Setting of auroral electron (a) energy and (b) number flux for test the polar auroral variation and (c) estimated UV and IR intensity variations for constant energy with variable flux (solid) and for variable energy with constant flux (dot-dashed and dotted) cases.

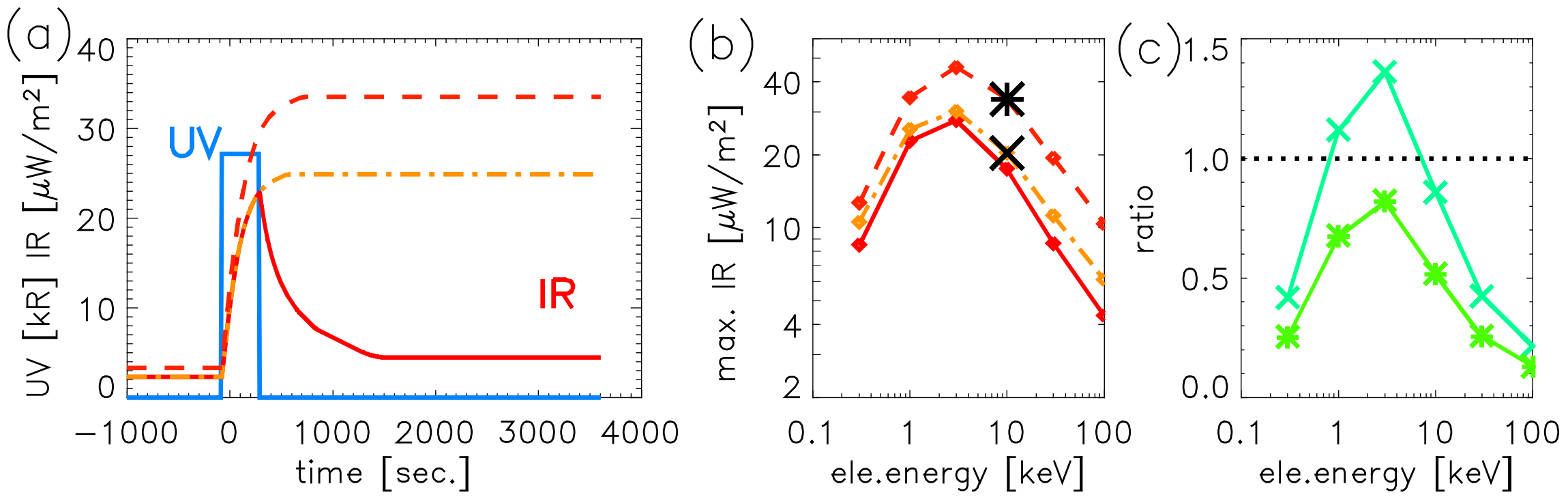

Fig.8 (a) Time variation of UV (blue/gray line) and IR (red/black) intensity for the 10 keV electron energy case and (b) maximum IR intensity as a function of electron energy. The cases of exospheric temperature and electron precipitation time duration of “1000 K, 404-sec”, “1000 K, >3600 sec”, and “1200 K, >3600 sec” are shown by solid, dot-dashed, and dashed lines, respectively. (c) The ratio of IR intensity “1000 K, 404-sec” (solid line in (b)) to "1000 K, >3600 sec, 10 keV" (cross in (b)) and the ratio of “1000 K, 404-sec” (solid line in (b)) to"1200 K, >3600 sec, 10 keV" (asterisk in (b)) are shown by crosses and asterisks, respectively.